\documentclass[12pt,a4paper]{iopart}
\usepackage{iopams}
\usepackage{epsf}

\usepackage{graphicx}
\usepackage{dcolumn}
\usepackage{bbm}
\usepackage{multirow}

\begin{document}

\title{Bifurcations in the Lozi map}

\author{V Botella-Soler\dag , J M Castelo\ddag , J A Oteo\S\ and J
Ros\dag\ \footnote[4]{To whom correspondence should be addressed} }

\address{\dag Departament de F\'{\i}sica Te\`{o}rica  and IFIC,
Universitat de Val\`{e}ncia-CSIC,
46100-Burjassot, Val\`{e}ncia, Spain}

\address{\ddag\ Departament de F\'{\i}sica Aplicada, Universitat de
Val\`{e}ncia,  46100-Burjassot, Val\`{e}ncia, Spain}

\address{\S\ Departament de F\'{\i}sica Te\`{o}rica, Universitat de
Val\`{e}ncia,  46100-Burjassot, Val\`{e}ncia, Spain}

\ead{vicente.botella@uv.es, oteo@uv.es, rosj@uv.es}

\begin{abstract}
{We study the presence in the Lozi map of a type of
abrupt order-to-order and order-to-chaos transitions which are
mediated by an attractor made of a continuum of neutrally stable
limit cycles, all with the same period.}
\end{abstract}

\pacs {05.45.Pq, 05.45 Ac, 02.30 Oz}

\ams{37G35, 37M05, 74H99, 37D45}

\submitto{\JPA }
\today



\maketitle

\section{Introduction\label{sec:intro}}
In 1978, Lozi introduced in a short note \cite{Lozi} a
two-dimensional map the equations and attractors of which resemble those of
the celebrated H\'enon map \cite{henon}. Simply, a quadratic term in
the latter is replaced with a piecewise linear contribution in the
former. This  allows one to rigorously prove the chaotic character
of some attractors \cite{Mcz} and a detailed analysis of their basins
of attraction \cite{Bap}. The equations for the iterated Lozi map
$L(x,y)$ read
\begin{eqnarray}\label{eq:Lozi}
x_{n+1}&=& 1+y_n-a|x_n| \equiv f(x_n,y_n) \nonumber\\
y_{n+1}&=& b x_n \equiv g(x_n,y_n),
\end{eqnarray}
where $a,b$ are real non-vanishing parameters. Inside the
region where the orbits remain bounded, the Lozi map may present both
regular and chaotic behavior.

The purpose of this article is to show that the Lozi map also exhibits a
special type of bifurcation recently found in a class of one
dimensional discontinuous piecewise linear  maps \cite{VGH}. A
similar phenomenon in continuous maps was also described in
\cite{Gardini1}. The main feature of this bifurcation is the
presence of a continuum of neutrally stable cycles. Thus, at the
very bifurcation point, the attractor in phase space contains an
infinite set of regular orbits. As a function of the parameters the
transit may take place from regular to regular as well as to chaotic
regimes.

In maps in one dimension this kind of transitions happens whenever a
piece of an iterate of the map becomes co-linear with the identity
function, the bisectrix, in the cobweb for a particular value of the parameter. We
will extend this explanation to maps in two dimensions and discuss
its connection with the theory of border collisions, which is
commonly used to analyse these kinds of maps \cite{NY,BG,diB}. Very recently, this
question has been considered from a more abstract point of view
\cite{Gardini2} and the name \emph{degenerate bifurcation} has been
used. As this term may sound polysemous we refer to this
phenomenon as \emph{bisecting bifurcation} in the rest of the paper,
which seems more in accordance with the geometrical interpretation
advanced.

An overview of the results on the Lozi map necessary for later
reference is presented in section \ref{sec:Bifs}. Bisecting
bifurcations are  introduced and analysed in section
\ref{sec:bisect}. They are then  given a geometrical explanation in
section \ref{sec:Explica}. In section \ref{sec:Ilustra} we further
illustrate how the dynamics of the map changes with parameter $a$
and establish a connection with the theory of border collisions.
Section \ref{sec:Fin} contains some final comments on somewhat
related topics.

\section{Dynamics of the Lozi map: isolated attractors and stability} \label{sec:Bifs}

It is customary to start the study of a dynamical system with a
catalogue of its fixed points, periodic attractors and so on, paying
special attention to their stability and evolution with the system
parameters. We start by collecting some of these features for the
Lozi map which will be referred to in the following sections.

For stability considerations we need the Jacobian matrix of $L(x,y)$
which reads
\begin{equation}\label{eq:Jacobian}
J(x,y)=\left(
\begin{array}{cc}
-a\,\mathrm{sign} (x)&1\\
b&0
\end{array}
\right) .
\end{equation}
Notice that $J(x,y)$ depends on the point of the orbit only through
$\mathrm{sign} (x)$. Accordingly, we denote its values as $J_+\equiv
J(x>0,y)$ and $J_-\equiv J(x<0,y)$. Furthermore, since
$\mathrm{det}\, J_{\pm}=-b$, we will only consider maps with $|b|\le
1$, in order to have non-expanding systems. In particular, the maps
with $b=\pm 1$ are area-preserving.

It is a simple exercise to see that for $|b|\le 1$ the Lozi map has
no fixed points if $a \le b-1$. It has $P_1$ as the unique fixed
point if $b-1 < a \le 1-b$, and gains an additional fixed point,
$P_2$, if $a>1-b$, with
\begin{equation}\label{eq:pfix}
\fl P_1(a,b)\equiv\Big(\frac{1}{1+a-b},\frac{b}{1+a-b}\Big)\qquad
P_2(a,b) \equiv \Big(\frac{1}{1-a-b},\frac{b}{1-a-b}\Big).
\end{equation}
Furthermore $J(P_1)=J_{+}, \, J(P_2)=J_{-}$. Only $P_1$ may be stable
and this happens for systems with parameter values $(b,a)$ inside the
triangle with vertices $(1,0),(-1,2), (-1,-2)$ in parameter space.

As for isolated cycles of period $T=2$, it is easy to see that they exist, if
we keep $|b|<1$, for $a>1-b$. Their elements are $(z_1,bz_2)$ and
$(z_2,bz_1)$, with
\begin{equation}
    z_1=\frac{1-b+a}{(1-b)^2+a^2} \qquad z_2=\frac{1-b-a}{(1-b)^2+a^2}.
\end{equation}
They are stable for parameter values $(b,a)$ inside the triangle with
vertices $(0,1),(1,0), (1,2)$ in parameter space.

The  analysis so far is standard and may be pursued for more
complicated attractors. We have recalled the previous facts just to
better understand the  bifurcations which will be studied in the next section.

\section{Bisecting bifurcations and attracting sets}\label{sec:bisect}

The fixed points and cycles mentioned above were isolated points in the phase space. However, for particular values of the parameters $a$ and $b$ we can also find continuum sets of periodic points  which act as attractors for certain regions of phase space. We commence by explaining the approach we follow to locate algebraically which values of $a$ and $b$  give
rise to such infinite sets of simultaneous cycles. We solve first the
cases of periods two and four which we deem illustrate sufficiently the
procedure. After that, we completely determine the cycles' elements. Period three and five are then analyzed in the
same way and shown to produce a different pattern. The stability of the cycles is also resolved. Higher order
periodic attractors may be studied along the same line, of course at
the price of more involved algebra.

\subsection{Continuum of period$-2$ attractors}\label{sec:T2}

Let $\{(x_1,y_1), \, (x_2,y_2)\}$ be a period$-2$ cycle
of the Lozi map. Next,
instead of (\ref{eq:Lozi}) we use the equivalent second order
difference equation
\begin{equation}
x_{n+1}=1-a|x_n|+bx_{n-1},
\end{equation}
with the obvious corresponding change in the initialization of the iteration.
The periodicity condition for this system conveys
\begin{eqnarray}\label{eq:T2}
x_2&=&1-a|x_1|+bx_2, \\
x_1&=&1-a|x_2|+bx_1.
\end{eqnarray}
The corresponding $2-$cycles of the Lozi map will be $\{ (x_1,bx_2),
(x_2,bx_1)\}$.

When the system of equations above is compatible but indeterminate
then there exists an infinity of solutions and hence of period$-2$ cycles.
From an algebraic point of view this means that the coefficient
and the augmented matrices must both have rank one:
\begin{equation}
\fl
\left|
\begin{array}{cc}
 a\; \mathrm{sign}(x_1) & 1-b \\
 1-b & a\; \mathrm{sign} (x_2)
\end{array}
\right| =0, \quad \mathrm{rank }\,\left(
\begin{array}{ccc}
 a\; \mathrm{sign}(x_1) & 1-b & 1 \\
 1-b & a\; \mathrm{sign} (x_2) & 1
\end{array}
 \right) =1,
\end{equation}
which yield the constraints: $a>0$ and either $b+a=1$ or $b-a=1$. Correspondingly there are two one-parameter families of $2-$cycles in
$\mathbb{R}^2$, $\{ S_1,S_2 \}$ and $\{ U_1,U_2 \}$ with
\begin{equation}
\fl
S_1= (x,b(1-ax)/a), \quad  S_2= ((1-ax)/a,bx),
\quad (0<x<1/a,\, b=1-a)
\end{equation}
\begin{equation}
\fl
U_1= (x,-b(1+ax)/a),\quad U_2=(-(1+ax)/a,bx),
\quad (-1/a<x<0, \, b=1+a).
\end{equation}

For the $S-$family we have $J(S_1)J(S_2)=J_+^2$, whose
eigenvalues are $(a^2+2b\pm |a|\sqrt{a^2+4b})/2$. In the critical
case, $a+b=1$, they reduce to the values $1$ and $(1-a)^2$, and
hence these orbits are neutrally stable whenever $0<a<2$. Then the continuum they form is an attractor in phase space.

For the $U-$family we have $J(U_1)J(U_2)=J_-^2$, whose eigenvalues in the critical case, $b-a=1$,  are $1$
and $(1+a)^2>1$, with $a>0$. Therefore these orbits are unstable.

\subsection{Continuum of period$-4$ attractors}\label{sec:T4}

The search of sets of period$-4$ orbits yields the
following results. First, the constraints on the parameters are
$a=1+b$, and $0<b<1$, $1<a<2$,
for the rank of the corresponding matrices to be
equal to three,
allowing in this way the existence of a one-parameter family of solutions.
Second, if $(x_i,y_i)$ denotes the cycle
elements,  with $i=1,\ldots, 4$, then
\begin{eqnarray}\label{eq:c4}
\fl
x_1&=&\mu+2b\beta,\quad x_2=\mu -(b-1)\beta, \quad x_3=-\mu,
\quad x_4=-\mu+(b+1)\beta ,\\
\fl
y_1&=&bx_4,\quad  y_2=bx_1,\quad  y_3=bx_2,\quad  y_4=bx_3,
\end{eqnarray}
where we have used the real parameter $0<\mu <1$ and the definition
$\beta\equiv 1/(1+b^2)$.

One can readily deduce from the expression of the cycle elements
(\ref{eq:c4}) that the $x-$components are alternatively positive and negative. Therefore the corresponding Jacobian matrix reads
simply $(J_+J_-)^2$, whose eigenvalues are $1$ and $(1-a)^4$ in the
critical case. Hence, these period$-4$ orbits are neutrally stable and their union acts also as an attractor.

\subsection{Period$-3$ attractors}\label{sec:T3}

Following the same algebraic procedure as in the preceding
subsections, we have determined that an infinite set of cycles of
three elements exists only if $a=1$ and $b=-1$. However, in this
case the rank of the corresponding coefficient and augmented
matrices is equal to one and, as a consequence, we get a
continuum of solutions depending on two parameters.

A phase portrait which includes the period$-3$ orbits is shown in
Figure \ref{fig:T3}. Orbits of period$-3$ fill completely the white
triangle at the center of Figure \ref{fig:T3}, with right angle
vertex at $(0,0)$. The acute angle vertices are located at $(1,0)$
and $(0,1)$. The elements of the cycle are given by $(\xi,-\eta)$,
$(1-\xi-\eta,-\xi)$ and $(\eta,-1+\xi+\eta)$, in terms of the
parameters $0<\xi,\eta<1$. The remaining white areas correspond to
regular orbits of higher periods. Finally, the black and grey areas
stand for chaotic trajectories.

The two eigenvalues of the Jacobian matrix of $L^{[3]}(x,y)$
evaluated on the orbit elements are equal to one, which stands for
double neutral stability. Hence,
the set of orbits of period$-3$ is two-dimensional in phase space,
at variance with period$-2$ and $-4$, and is embedded into the sea of
chaotic trajectories.

\subsection{Period$-5$ attractors}\label{sec:T5}

The situation with period$-5$ attractors is similar to the one with
period$-3$. We have found two two-parameter sets of orbits with
neutral stability. They emerge when $b=-1$ and either $a=(1+\sqrt
5)/2\equiv \phi$ (the golden ratio) or $a=(1-\sqrt 5)/2$. Table
\ref{Table-T5} collects the elements $\{x_i,y_i\}$, with
$i=1,\ldots,5$, of the orbit. We use the notation $\varphi \equiv
(\sqrt 5 -1)/2 =\phi -1=1/\phi$. The two parameters $u$ and $v$
which define the family of period$-5$ orbits vary in the pentagon of
vertices
$\{(\varphi^2,0),(\varphi,0),(\varphi^2,\varphi^2),(0,\varphi),(0,\varphi^2)\}$
in parameter space,
provided $a=\phi$. Else, if $a=-\varphi$, the two parameters vary in
the pentagon of vertices $\{(0,0),(1,0),(\phi,1),(1,\phi),(0,1)\}$. As far as stability is concerned, a direct calculation shows that the Jacobian matrix of $L^{[5]}(x,y)$ for the parameter values just quoted
is indeed the unit matrix which means double neutral stability.
\begin{table}[hb]
\begin{center}
      \begin{tabular}{lcccccr}
        \hline\hline
         $a$&$i$& 1 & 2 & 3 & 4 & 5 \\
         \hline
        \multirow{2}{*}{$\phi$}&$x_i$ & $1-\phi v-u$ & $-\varphi +\phi (u+v)$& $1-\phi u-v$&$u$ & $v$ \\
        &$y_i$ & $-v$ & $\phi v+u-1$ & $\varphi -\phi (u+v)$ & $\phi u+v-1$ & $-u $\\
        \hline
        \multirow{2}{*}{$- \varphi$}&$x_i$ & $1+\varphi v-u$ & $\phi -\varphi (u+v)$& $1+\varphi u-v$&$u$ & $v$ \\
        &$y_i$ & $-v$ & $u-\varphi v-1$ &  $\varphi (u+v) -\phi$& $v-\varphi u-1$ & $-u$ \\
        \hline\hline
      \end{tabular}
\caption{Elements of the period$-5$ orbits $\{x_i,y_i\}$,
$(i=1,\ldots,5)$. The upper block entry is for $a=\phi$. The real
parameters $u,v$ take values inside and on the pentagon of vertices
$\{(\varphi^2,0),(\varphi,0),(\varphi^2,\varphi^2),(0,\varphi),(0,\varphi^2)\}$
of parameter space.
The lower block entry is for $a=-\varphi$ and the parameters are
inside and on the pentagon of vertices
$\{(0,0),(1,0),(\phi,1),(1,\phi),(0,1)\}$ in parameter space.
In both cases it is $b=-1$. All
the orbits elements stay in the first quadrant of phase plane.}
\label{Table-T5}
\end{center}
\end{table}

\subsection{Bifurcation diagrams}

It is customary to summarize  the behavior of a map with varying parameters by means of bifurcation diagrams. In principle there are two somewhat extreme ways to construct these diagrams. The ideal mode to proceed is clear: one  determines by a theoretical analysis the periodic points and their stability and plot them as functions of a parameter. This is often impossible to carry out completely and then one has to resort to  a more {\it experimental} way: iterate the map for a number of initial conditions, discard a transient in the trajectories and plot then a component as a function of the parameter. The unavoidable limitations of the last approach are clear and one has to take special care to identify the results obtained. In particular, a densely filled bar in a diagram is usually taken as a signature of chaos.

\begin{figure}[H]
\vspace{2cm}
\includegraphics[scale=0.3]{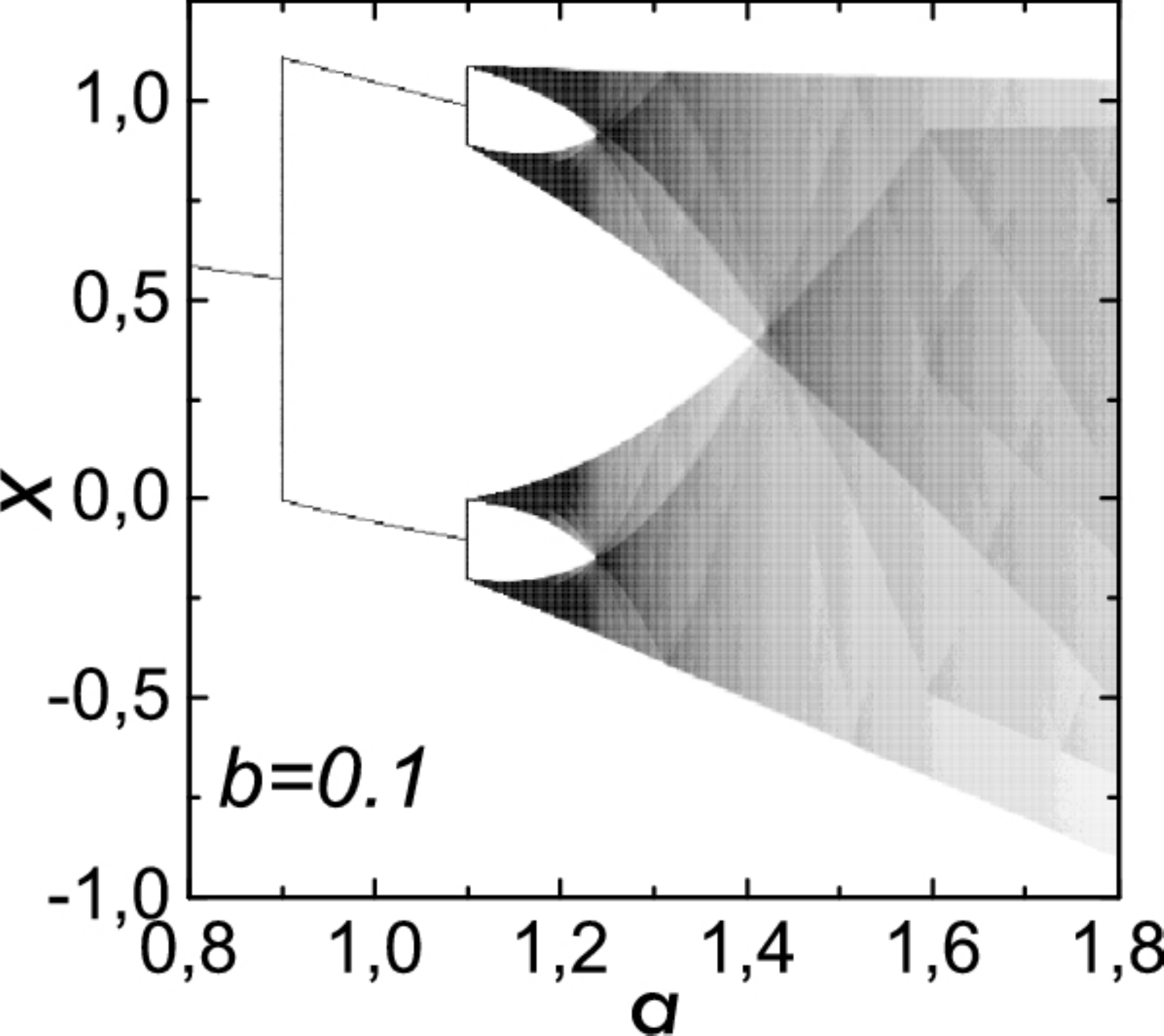}
\caption{\label{fig:Bif-a} Bifurcation diagram for $b=0.1$. Notice
that the bisecting bifurcations take place at $a=1\pm b $.}

\vspace{2cm}
\includegraphics[scale=0.3]{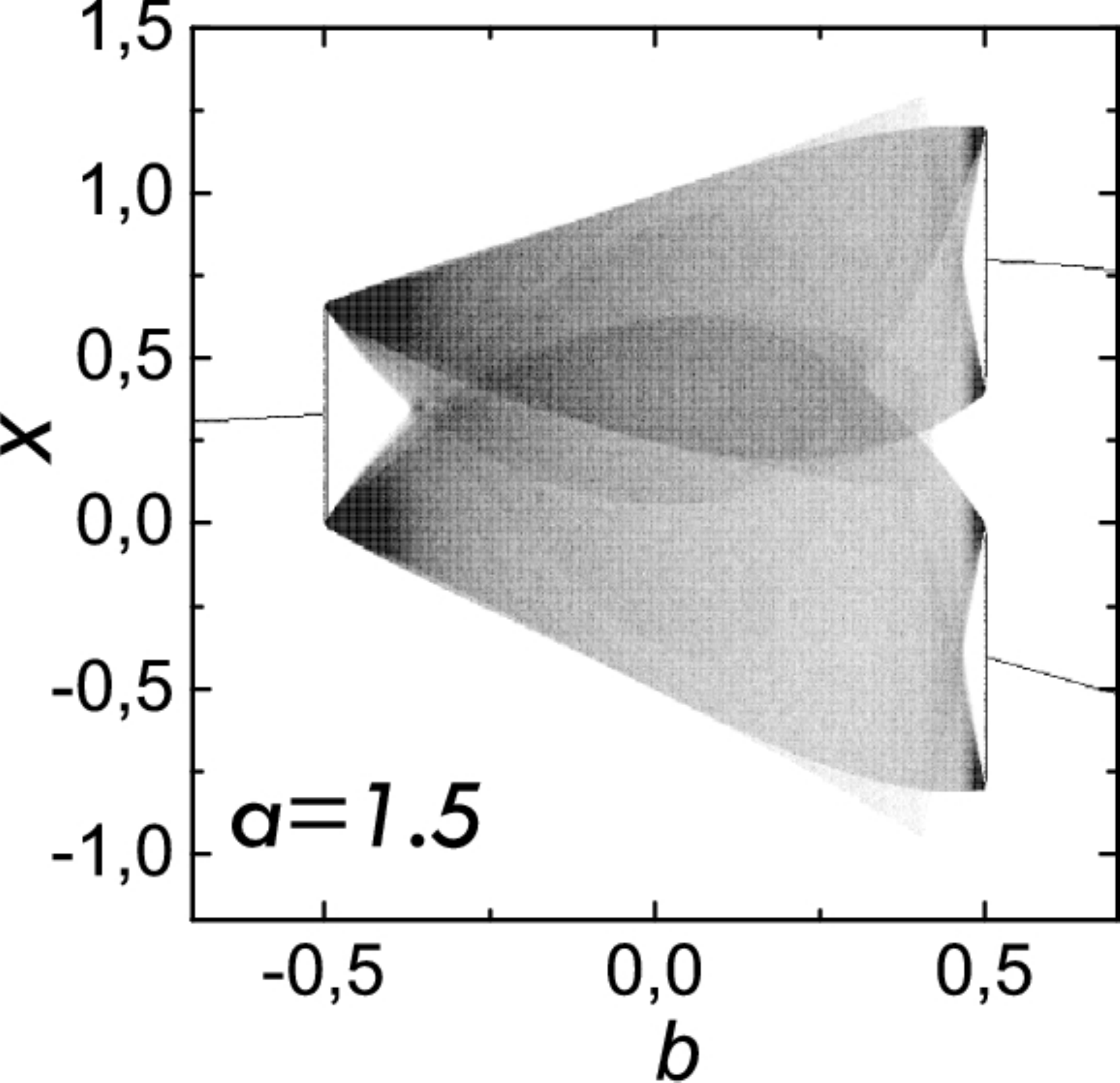}
\caption{\label{fig:Bif-b} Bifurcation diagram for $a=1.5$. Notice
that the bisecting bifurcations take place at $b=\pm (a-1)$.}
\end{figure}

In the Lozi map, when enough initial conditions are considered for the numerical simulation, the continua of $2-$ and $4-$cycles discussed above may act as an attractor and show up also as a vertical line in the bifurcation diagram. This is clearly seen in Figure \ref{fig:Bif-a} for fixed
$b=0.1$. To construct this diagram especial care has been payed to sweep a large number of initial conditions.  For every trajectory a large enough transient is discarded. Also, if a component increases beyond a given threshold  the trajectory is considered as unbounded. Figure \ref{fig:Bif-a} exhibits a bifurcation from period$-1$ to period$-2$ at
$a=0.9$ and another one from period$-2$ to chaos at $a=1.1$. This numerically obtained diagram illustrates clearly that these two bifurcations are mediated by the phenomena described in sections \ref{sec:T2} and \ref{sec:T4} respectively. A further feature of Figure \ref{fig:Bif-a} is that in the approximate intervals
$1.1<a<1.237$ and $1.237<a<1.406$, the trajectories jumps between the chaotic bands in a cyclic manner,
with periods 4 and 2 respectively.

Similarly, in Figure \ref{fig:Bif-b}, here for fixed $a=1.5$, two
bifurcations occur at $b=\pm 0.5$, again mediated by the same mechanism.
The system is chaotic for $b=0.1$ and $a>1.1$
in Figure \ref{fig:Bif-a}, and $a=1.5,\, -0.5<b<0.5$ in Figure
\ref{fig:Bif-b}. The presence of these vertical
segments corresponding to the described attractors composed of
infinite sets of periodic orbits is the most salient feature
of the phenomenon we deal with here. We term it {\it bisecting
bifurcations} for reasons which will appear clear in the next section.
Eventually, cyclic jumping between the chaotic bands of the diagram takes place
with period 2 and 4 to the left and right of the figure, respectively.

We gather the information discussed so far in Figure \ref{fig:Lyap}
where the behavior of the Lozi map in different regions of the parameter space is indicated.
Trajectories
for $|b|>1$ are unbounded.  Numerical simulations indicate this is
also the case for  the northern  and southern parts  of the diagram.
The identification of the chaotic area has been achieved by computing numerically
the Lyapunov exponents \cite{Yorke}, thus the upper border of chaos
is approximate. The stable fixed points of $L$ and $L^{[2]}$,
are included in the graphics.  The borders of their stability region will be
analysed in detail later on. The diagram by no means exhausts the
variety of behaviors the trajectories of the Lozi map may develop
\cite{Elhadj}.

\section{A geometrical explanation for the bisecting
bifurcations in the Lozi map}\label{sec:Explica}

\begin{figure}[H]
\includegraphics[scale=0.5]{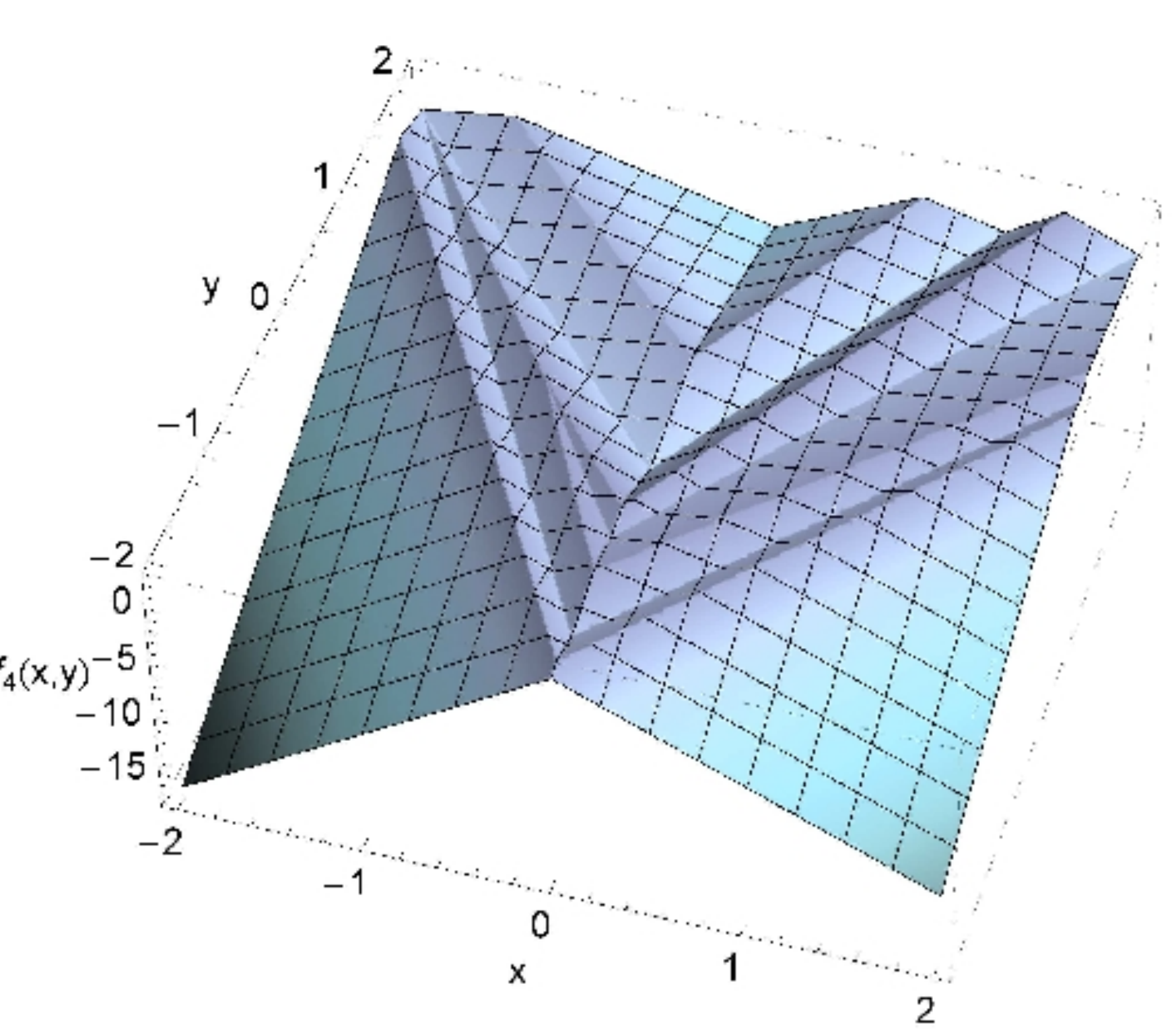}
\caption{\label{fig:Lozif4} (Color online) Surface defined by the iterate $f^{[4]}(x,y)$
in the definition of the Lozi
map, with $a=1.5$ and $b=0.5$.}

\vspace{2cm}
\includegraphics[scale=0.6]{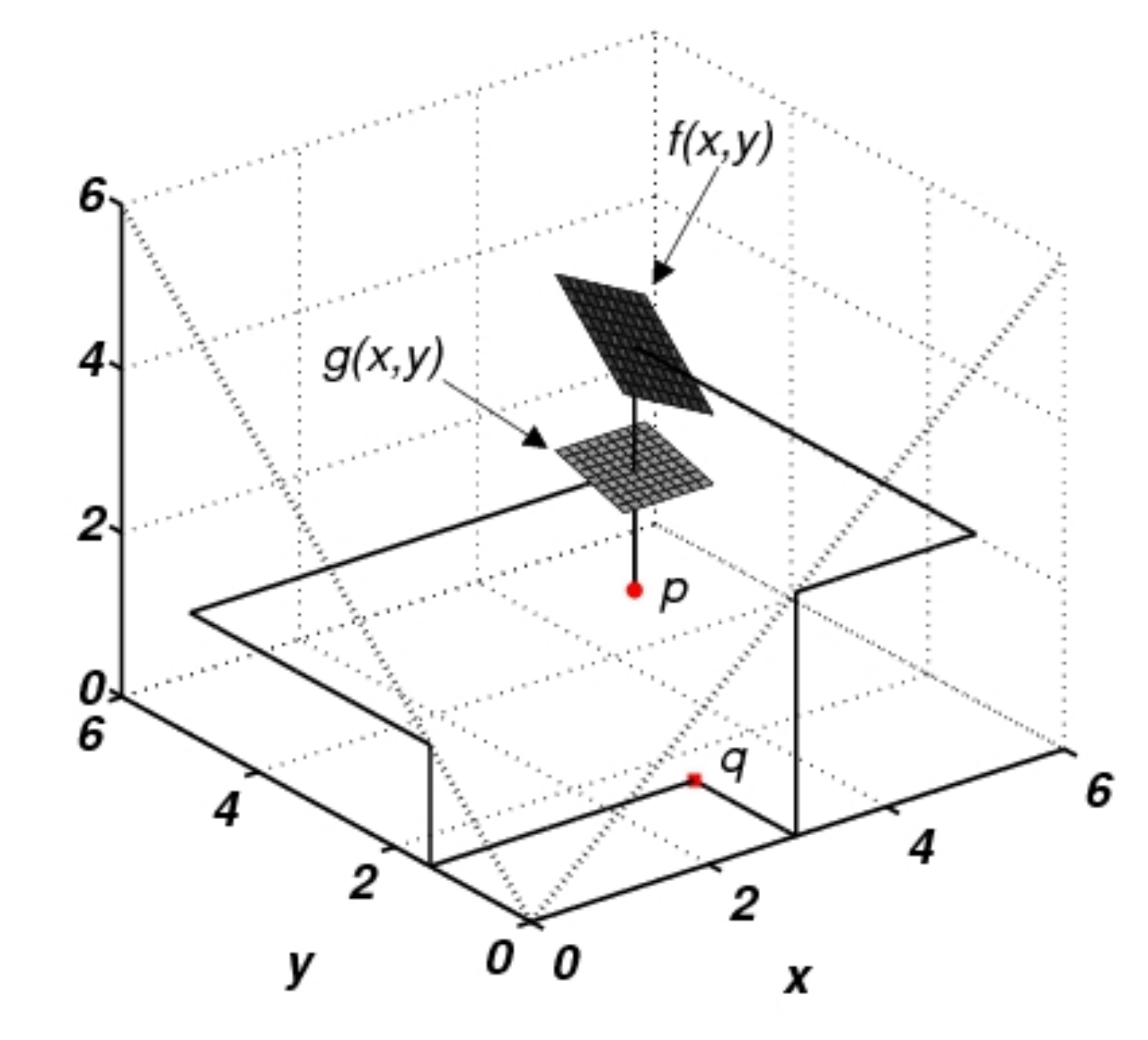}
\caption{\label{fig:cobweb} (Color online) Example of two-dimensional cobweb
diagram for the Lozi map ($a=0.6$, $b=0.3$). The point
$p\in\mathbb{R}^{2}$ is projected to the surfaces $f(x,y)$ and
$g(x,y)$ and these projections are brought back down to
$\mathbb{R}^{2}$ by means of reflections in the bisectrix $f(x,0)=x$
and $g(0,y)=y$ in order to find the new point $q$. For the sake of
clarity only one small region of each surface is shown. }
\end{figure}

In \cite{VGH}, a geometrical explanation for the bisecting
bifurcations in piecewise one-dimensional maps is provided. Here we
develop a generalization for the two-dimensional Lozi map, although the
graphical illustration is more involved. Figure \ref{fig:Lozif4}
represents the surface $f^{[4]}(x,y)$ and allows us to appreciate
how the simple shape of $f(x,y)$ in (\ref{eq:Lozi}) becomes convoluted
after iteration.

Notice that for the Lozi map, the existence of a bisecting
bifurcation in the variable $x$ implies its presence in the variable
$y$, and viceversa. This is due to the fact that $y$ is just a one
iteration delayed and re-scaled version of $x$. This observation
will render easier the design and interpretation of some three
dimensional diagrams.

From an algebraic point of view, we will find an infinity of
period-$n$ limit cycles if
\begin{equation}\label{condLozi}
f^{[n]}(x,y)=x,\quad g^{[n]}(x,y)=y,
\end{equation}
for some range of $(x,y)$ values, as illustrated for $n=2$ to $5$ in
section \ref{sec:Bifs}. Clearly, this is the condition of fixed
point extended to a continuous set of points in phase space.
Furthermore, this is only a necessary condition for the observation
of the infinity of limit cycles since its stability is not assured
by it.

In one-dimensional maps it is a common practice to construct and
study trajectories using the cobweb diagram. In a two-dimensional
generalization we can visualize $f^{[n]}(x,y)$ and $g^{[n]}(x,y)$ as
surfaces to which a point $p\equiv(x_m,y_m)$ of the phase plane is
projected upward in order to find the new $x_{m+1}=f^{[n]}(x_m,y_m)$
and $y_{m+1}=g^{[n]}(x_m,y_m)$ on the vertical axis. This procedure
is illustrated in Figure \ref{fig:cobweb} for the particular case
$n=1$. The horizontal plane stands there for the phase plane. The
vertical axis is common for both surfaces $f^{[n]}(x,y)$ and
$g^{[n]}(x,y)$, which have been sketched only partially. Using the
cobweb diagram technique, the values $x_{m+1}$ and $y_{m+1}$ are
sent back to the phase plane. Thus, the value $x_{m+1}$ is first
projected toward the bisectrix line $f^{[n]}(x,0)=x$ and then down
to the $x$-axis. Similarly, $y_{m+1}$ is first projected toward the
bisectrix $g^{[n]}(0,y)=y$ and then down to the $y$-axis. This gives
the point $q\equiv(x_{m+1},y_{m+1})$ on the phase plane of Figure
\ref{fig:cobweb}.

Let us now interpret condition (\ref{condLozi}) in geometrical
terms. The presence in a two-dimensional map of a bifurcation
mediated by an infinite set of neutrally stable cycles implies the
existence of a segment, or set of segments, in the phase plane whose
projections upward on the surfaces $f^{[n]}(x,y)$ and $g^{[n]}(x,y)$
are contained in the bisecting planes $f^{[n]}(x,y)=x$ and
$g^{[n]}(x,y)=y$ respectively. This explanation justifies the name
\emph{bisecting bifurcation} we have given to the phenomenon we are
discussing.

In the case of the Lozi map, $f$ and $g$ allow explicit solutions
for condition (\ref{condLozi}) to be found, at least for low values
of $n$. Next we focus on the particular case of the bifurcation
occurring at $a=1-b$ with $a>0$. In this case the limit cycles at
the bifurcation point are all period$-2$ cycles, i.e. fixed points
of $L^{[2]}(x,y)$. In Figure \ref{fig:cobweb2} we have represented
the surface $f^{[2]}(x,y)$ and the generalized cobweb procedure
described above. For the sake of clarity, any analogous
representation concerning $g^{[2]}(x,y)$ has been omitted. The plot
has been generated for $b=0.1$ and $a=0.9$. Any initial condition
$(x_0,y_0)$ lands, after a certain transient, on the segment
$y=b(1-ax)/a$ with $0<x<1/a$ (red line in Figure \ref{fig:cobweb2}).
Every point of this segment is a fixed point of $L^{[2]}(x,y)$. When
projected to the surface $f^{[2]}(x,y)$ they graze the bisector
plane $f^{[2]}(x,y)=x$ and then come back to their position in the
phase plane. Since this segment is invariant under the
transformation $L^{[2]}(x,y)$, any point in the segment is an
element of a period$-2$ cycle of $L(x,y)$.

In summary, from a three-dimensional geometrical perspective, we
will find an infinite set of limit cycles of
period $n$ from the Lozi map when
the projections on the phase plane of the two curves produced by the
intersections of i) the surface $f^{[n]}(x,y)$ and the bisector
plane $z=x$, and ii) the surface $g^{[n]}(x,y)$ and the bisector
plane $z=y$, overlap in a certain range of the variables $x$,$y$.
Whenever the intersection is not a linear segment but an area, the
system has double neutrally stable orbits, which is the case of the
period$-3$ and $-5$ orbits we have studied.

\section{Bifurcations and phase space}\label{sec:Ilustra}

\begin{figure}[H]
\includegraphics[scale=0.6]{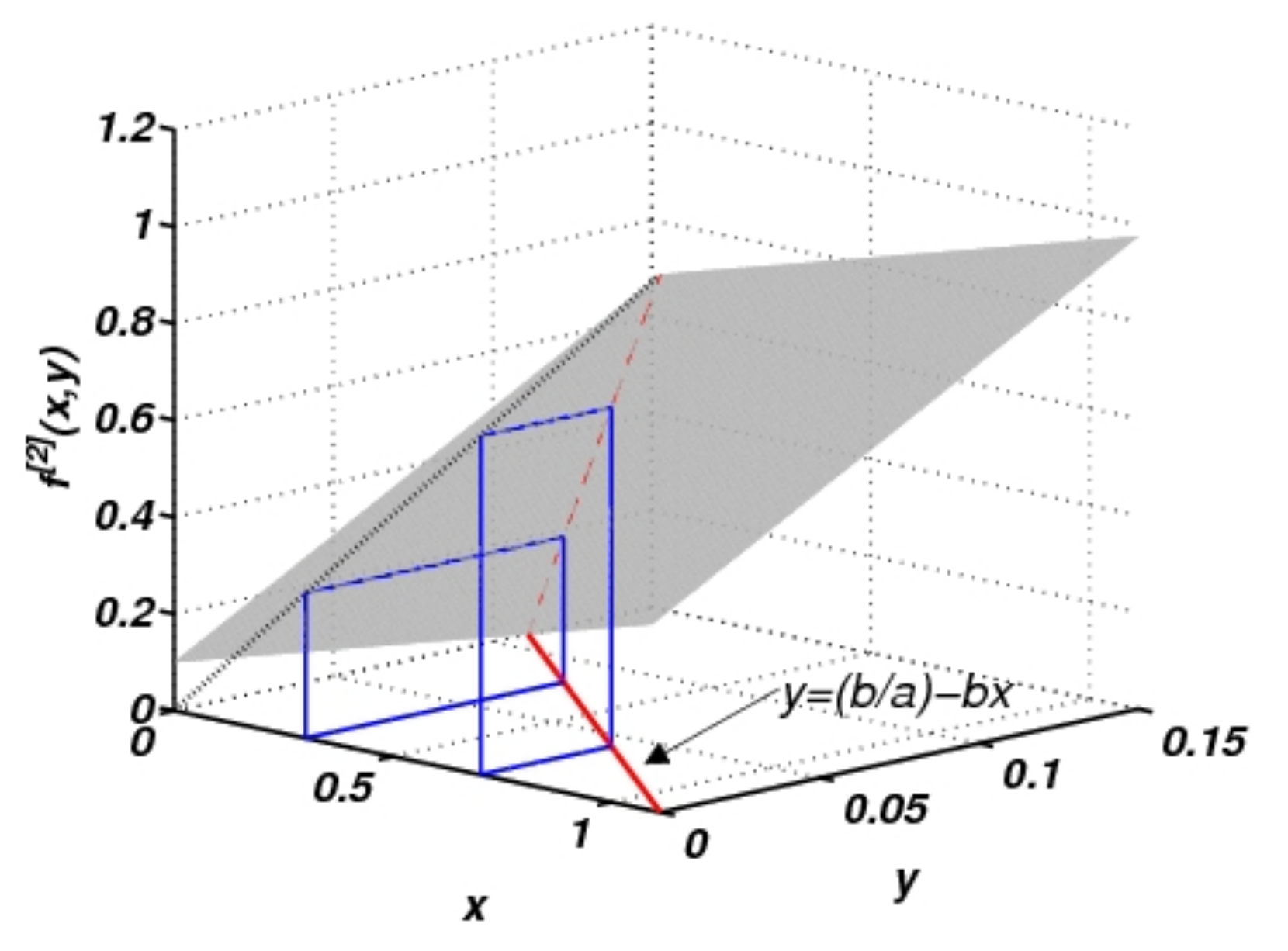}
\caption{\label{fig:cobweb2} (Color online) Cobweb diagram for the 2nd iterate of
the Lozi map with $b=0.1$ and $a=1-b$. Only the $f^{[2]}(x,y)$
surface is shown. The cobweb trajectories (blue) of two different
initial conditions ($x_{0}=0.3,0.7$ and $y_{0}=b(1-x_{0})/a$)
have been represented.}

\vspace{2cm}
\includegraphics[scale=0.4]{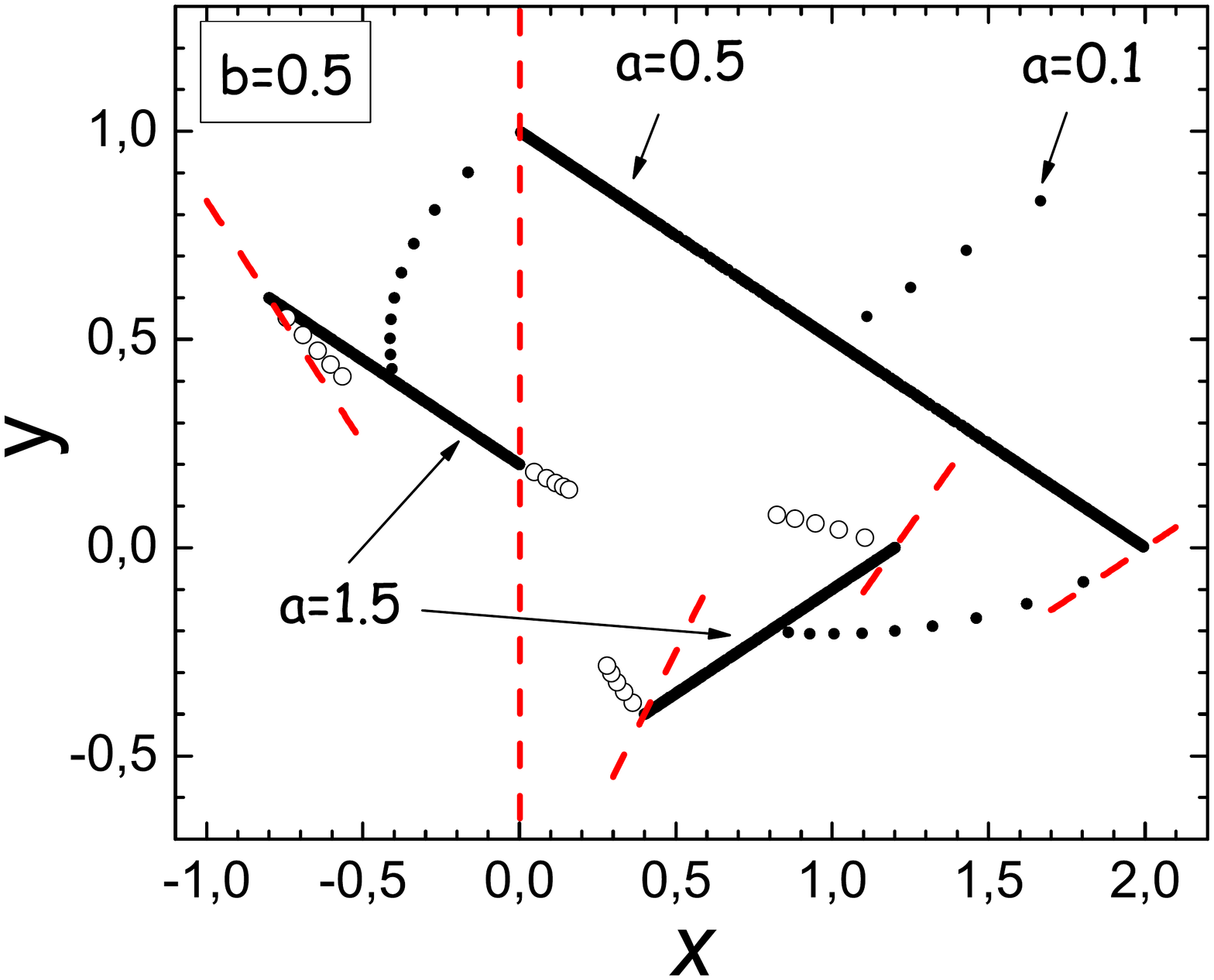}
\caption{\label{fig:attractor1} (Color online) Superposition of twenty attractors
illustrating border collisions. $b=0.5$ and $a=0.1\,n$ with
$n=1,2,\ldots,20$. Period$-2$ and period$-4$ attractors associated
to neutrally stable orbits ($a=0.5$ and $1.5$) correspond to the
segments (solid pattern). In between, stable period$-2$ limit
cycles. The dashed lines stand for the borders at the bifurcation
values (as a matter of fact, the border $x=0$ is invariant). The
circles stand for a period$-4$ unstable orbit plotted for the values
$a=0.1\,n$, $n=16,\ldots,20$ which collides with the borders at the
bifurcation.}
\end{figure}

In this section we discuss the appearance of the bisecting
bifurcations in connection with the changes in the stability of
other attractors. This will further clarify the meaning of the
vertical segments in bifurcation diagrams like the ones in Figures
\ref{fig:Bif-a} and \ref{fig:Bif-b}.  To be specific we fix $b=0.5$
and plot in Figure \ref{fig:attractor1}  twenty
regular attractors in phase space corresponding to parameter values $a=0.1\,n$,
$n=1,2,\ldots, 20$.

The diagram allows us to appreciate how the fixed point attractor
evolves with increasing $a $ for $n=1,\ldots,4$. At $a=1-b=0.5$,
i.e.  $n=5$, the first bisecting bifurcation takes place. Observe
that this is precisely where the neutrally stable fixed point
$P_1(a,b)$ in (\ref{eq:pfix}) changes to be unstable. We note in
passing that this type of behavior sounds familiar from the analysis
of elementary systems. For example, in the most conventional
period-doubling cascade that would mean the birth of a period$-2$
attractor. Here something of this sort is also observed because
$2-$cycles exist to the right of  $a=0.5$. The difference is that
the transition is mediated by a continuum of  neutrally stable
cycles of period$-2$ which fill up the segment in the figure.

When we continue to increase $a$  the system is periodic with
stable period$-2$ cycles. These have been represented in the figure for
nine values of $a$, namely  $n=6,\ldots,14$. Then, at $a=1+b=1.5$,
the second bisecting bifurcation takes place. Again these parameters
values punctuate the transition from stable to unstable for the
2-cycles mentioned in section \ref{sec:Bifs}. The two segments of
the attractor are now built up by neutrally stable cycles of
period$-4$, and they are apparent in Figure \ref{fig:attractor1}.
Values $a>1.5$ convey chaos (not shown) with unstable cycles in it
($n=16,\ldots,20$).

\begin{figure}[H]
\vspace{2cm}
\includegraphics[scale=0.4]{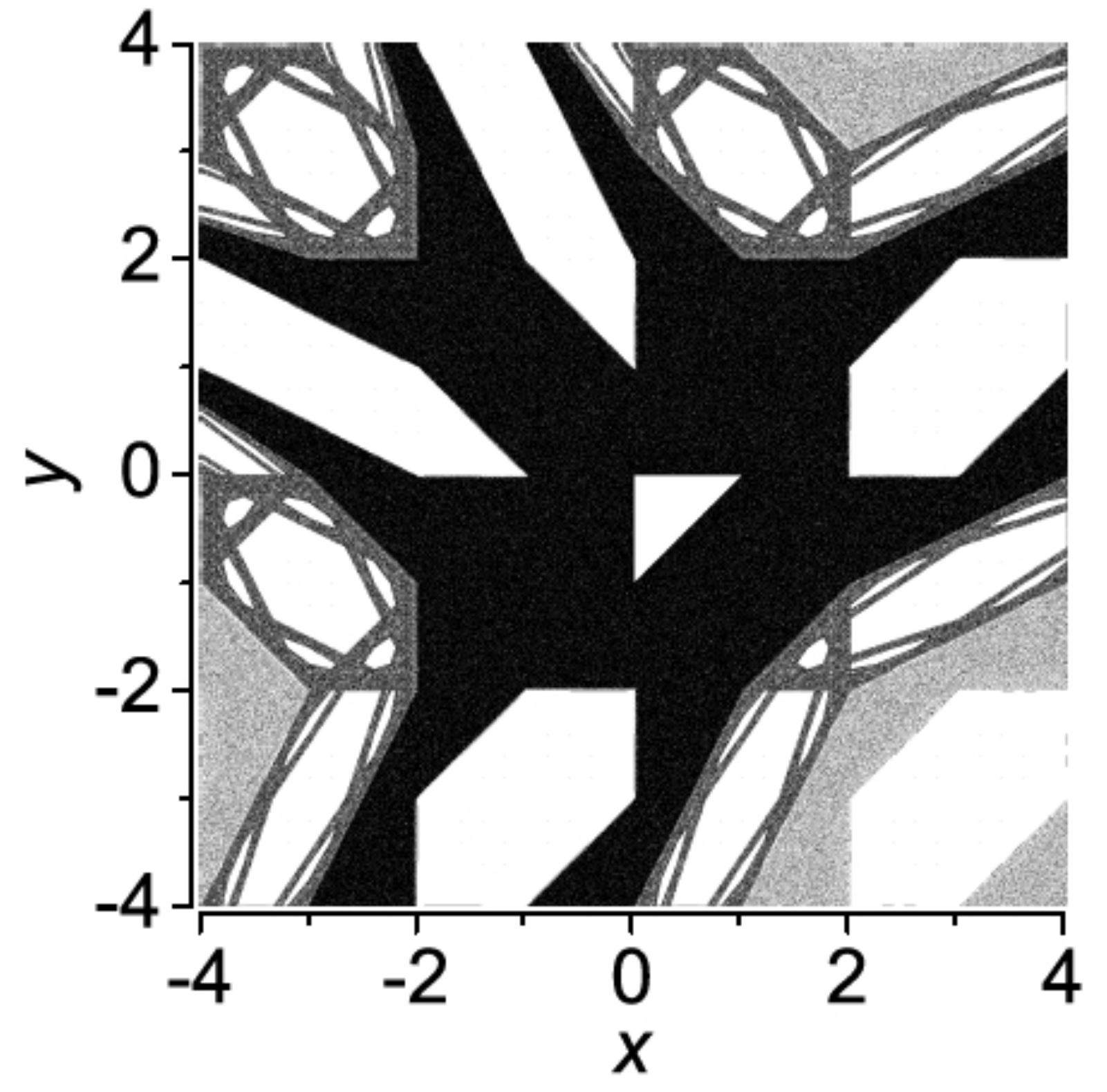}
\caption{\label{fig:T3} Attractor of the system $a=1$, $b=-1$.
Black and gray regions stand for chaotic trajectories.
The white areas contain only periodic orbits with neutral
stability. In particular, the innermost triangle with
right angle vertex at $(0,0)$
contains only period$-3$ trajectories. }

\vspace{2cm}
\includegraphics[scale=0.6]{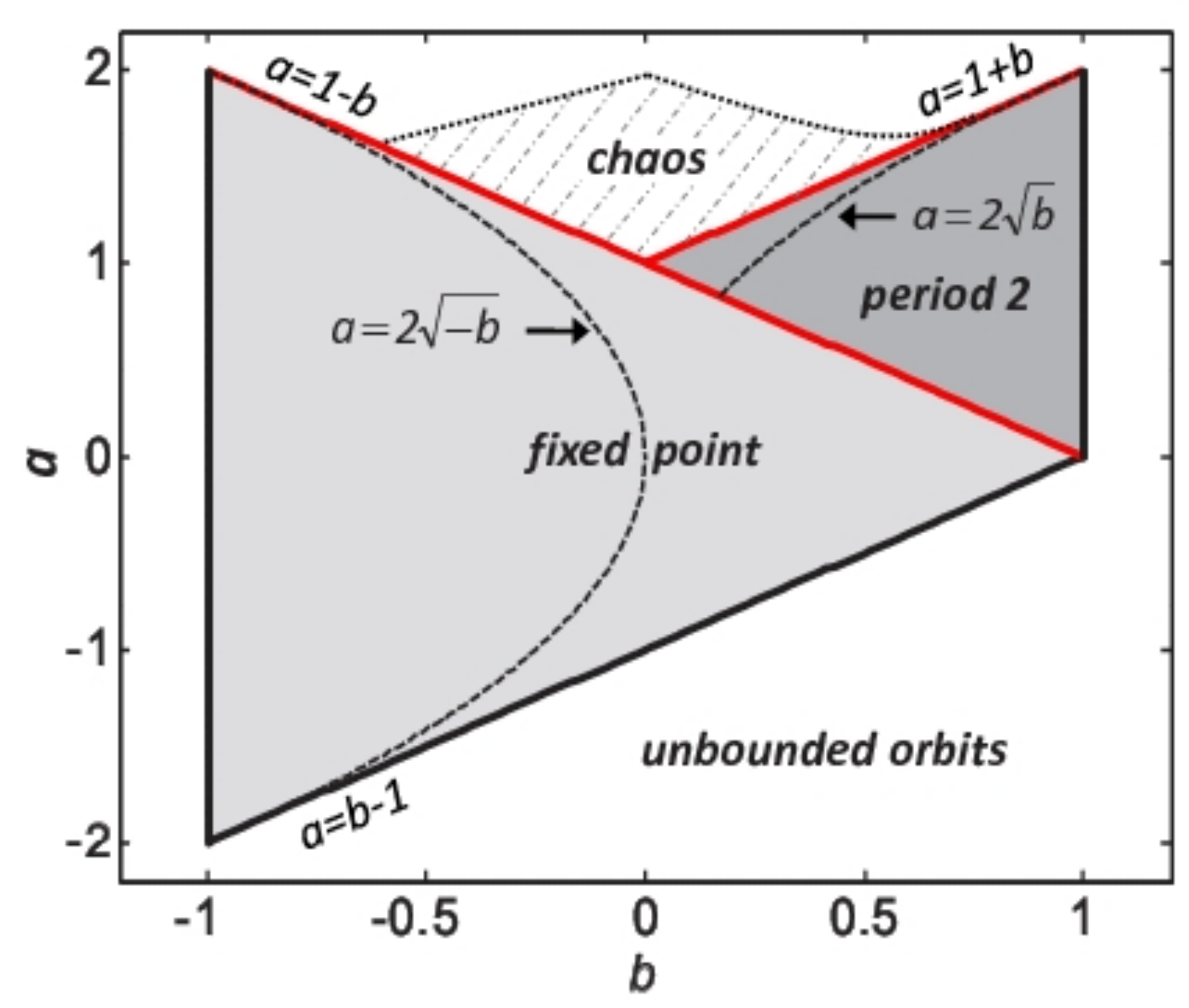}
\caption{\label{fig:Lyap} (Color online) Behaviour of the Lozi map
in parameter space. White areas correspond to unbounded
trajectories. The bisecting bifurcations take place at the red
straight lines, $a=1\pm b$. This diagram does not exhaust all the
possible attractors the Lozi map may exhibit.}
\end{figure}

The case considered illustrates, for fixed $b$ and varying $a$, how
the Lozi map follows a route to chaos in three steps separated by
two bifurcations  characterized by the coexistence of a
continuum of cycles with period $2$ and $4$ respectively. In Figure \ref{fig:Lyap} this
corresponds  to a raising vertical
trajectory in parameter space. As a matter of fact this figure
contains information on the whole family of bifurcation diagrams
like the ones in Figures \ref{fig:Bif-a} and \ref{fig:Bif-b} which
stem from, respectively, vertical or horizontal displacements in
parameter space. Accordingly, a variety of transitions are possible:
fixed point to chaos, periodic cycle to chaos or fixed point to
periodic cycle. We emphasize the role the bisecting bifurcations
play in mediating these transitions.

Since piecewise continuous maps are usually analyzed in terms of
border collisions  \cite{NY,BG}, it is appropriate to  discuss the
connection of the bisecting bifurcations  with that scheme. In the
particular case of the Lozi map, the described period$-2$ and
period$-4$ attractors involving neutrally stable cycles may be seen
as  border collisions too. To see how this comes about we propose to
look again at Figure \ref{fig:attractor1} but going in the reverse
direction. Starting with systems with $a=2$ and progressively
decreasing its value ($n=20,\ldots,16$), one finds unstable
period$-4$ orbits represented by circles. For $a=1.5$ these cycles
collide with the borders, namely the discontinuities of
$L^{[4]}(x,y)$ (dashed lines), originating the two segments made of
an infinite set of neutrally stable cycles of period$-4$. After the
collision, further decrease of $a$ gives then rise to stable
period$-2$ attractors for $a<1+b=1.5$. Following the way down, the
$2$-cycle collides, for $a=1-b=0.5$, with the discontinuities of
$L^{[2]}(x,y)$ and we see a sudden and ephemeral segment of
period$-2$ neutrally stable cycles. Finally, for $a<0.5$ we find the
stable fixed points $P_1$. Notice that in these collisions the
location of the borders depends on $a$, except the one at $x=0$
which is fixed.

\begin{figure}[H]
\vspace{1cm}
\includegraphics[scale=1.2]{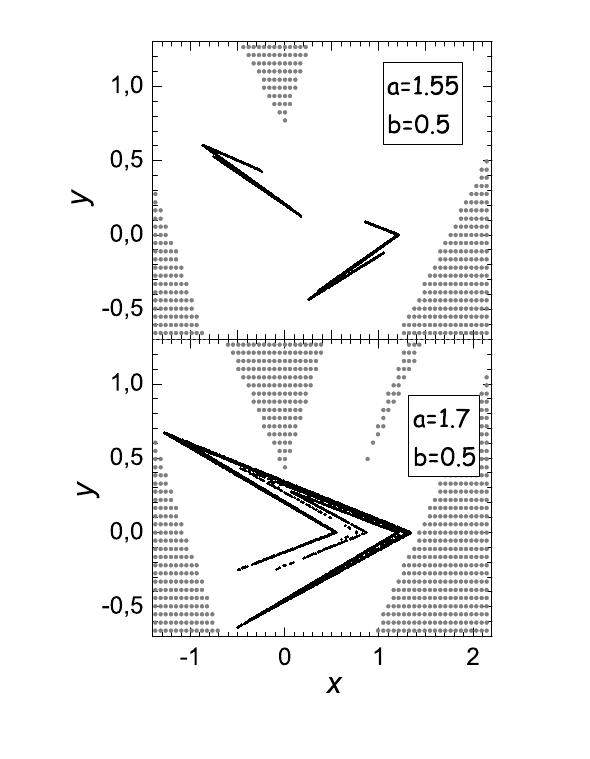}
\caption{\label{fig:attractor2} Two strange attractors of the Lozi
map for the values $a$ and $b$ specified in the panels. Scales are
the same as in Figure \ref{fig:attractor1} what allows to appreciate
how the regular attractor with $a=1.5$ in Figure
\ref{fig:attractor1} becomes strange as $a$ increases. Points in the dotted patterns give rise to unbounded orbits. }

\vspace{2cm}
\includegraphics[scale=0.5]{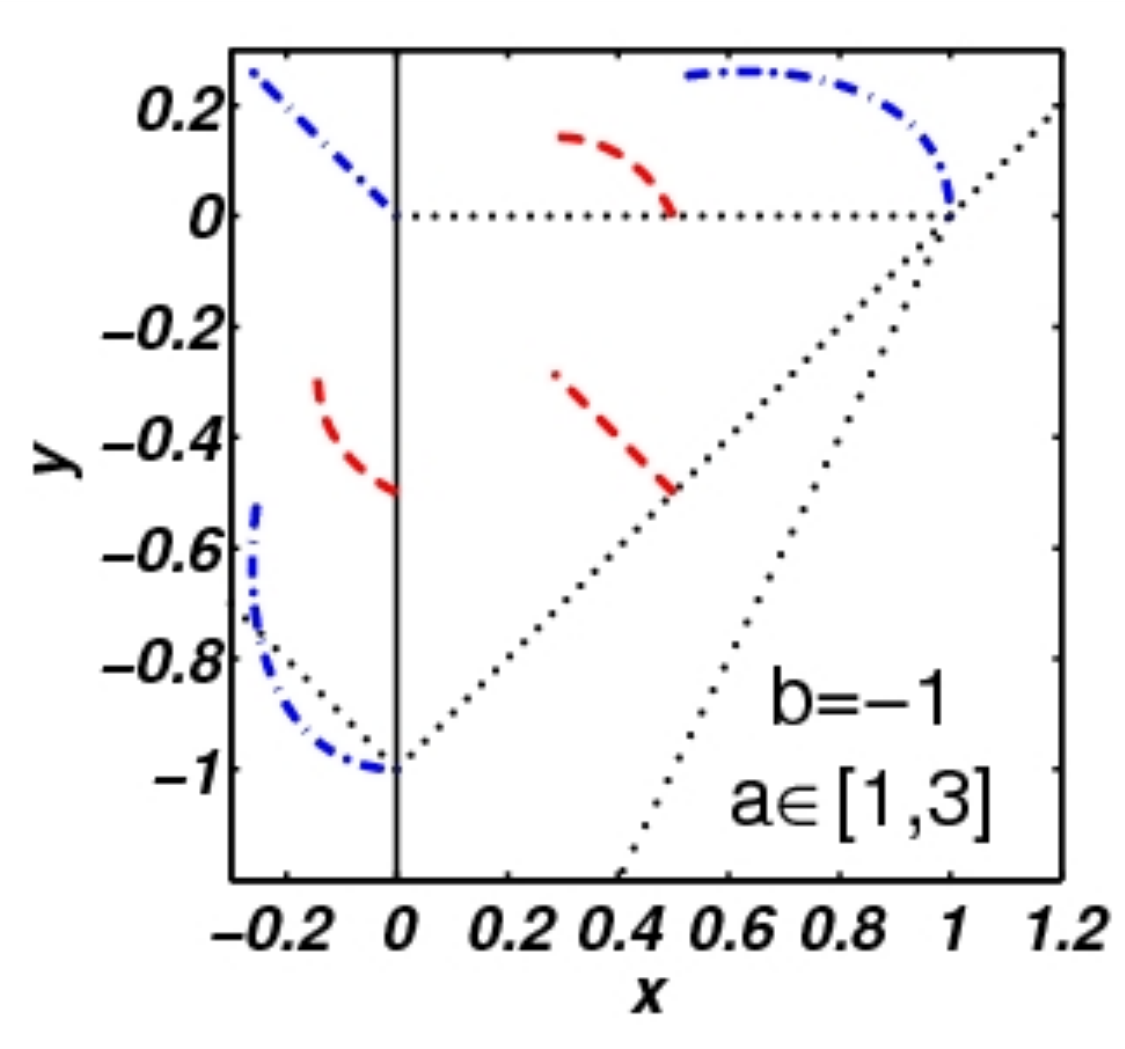}
\caption{\label{fig:attractorT3} (Color online) Period$-3$
border collision in the Lozi map, illustrated as a sequence of
snapshots in phase space. The solid line $x=0$ stands for a
border of the map, which is independent of $a$. The dotted
lines represent further borders of the map at $a=1, b=-1$. For
other values of $a\ne 1$ these lines wander across the plot.
The dashed and dot-dashed lines stand for the evolution of two
different orbits of period$-3$ as the value of $a$ varies in
the interval $[1,3]$. The border collision takes place at the
bifurcation values $a=-b=1$.}
\end{figure}

As regards the chaotic regime, two examples of attractors appear in
Figure \ref{fig:attractor2}, intended to be complementary of
Figure \ref{fig:attractor1}. The upper panel is for $a=1.55$ and
shows an incipient zigzagged structure in phase space as the result
of the destruction of the regular attractor $a=1.5$ in Figure
\ref{fig:attractor1}. In the bottom panel, with $a=1.7$, the strange
attractor is apparent. It corresponds to the figure in the original
reference \cite{Lozi}. This $V-$shaped structure seems to be the
prototype of chaotic attractor with $|b|<1$. Points in the dotted
pattern originate unbounded orbits.

So far in this section we have considered $|b|<1$. As has already
been mentioned, systems with $|b|=1$ are area-preserving and their
attractors offer a large variety of structures. Figure \ref{fig:T3}
is an instance of this. The system with $a=1,b=-1$, contains doubly
neutrally stable period$-3$ orbits embedded in chaos. In this regard,  Figure
\ref{fig:attractorT3} illustrates that  this situation corresponds also
to a border collision. It shows the evolution with $a$ of six
different fixed points of $L^{[3]}(x,y)$ (curves coded as dashed and
dot-dashed lines). The remaining lines in the plot stand for
borders. The collisions between fixed points and borders at $a=1$
are clearly observed.

\begin{figure}[H]
\vspace{1cm}
\includegraphics[scale=0.8]{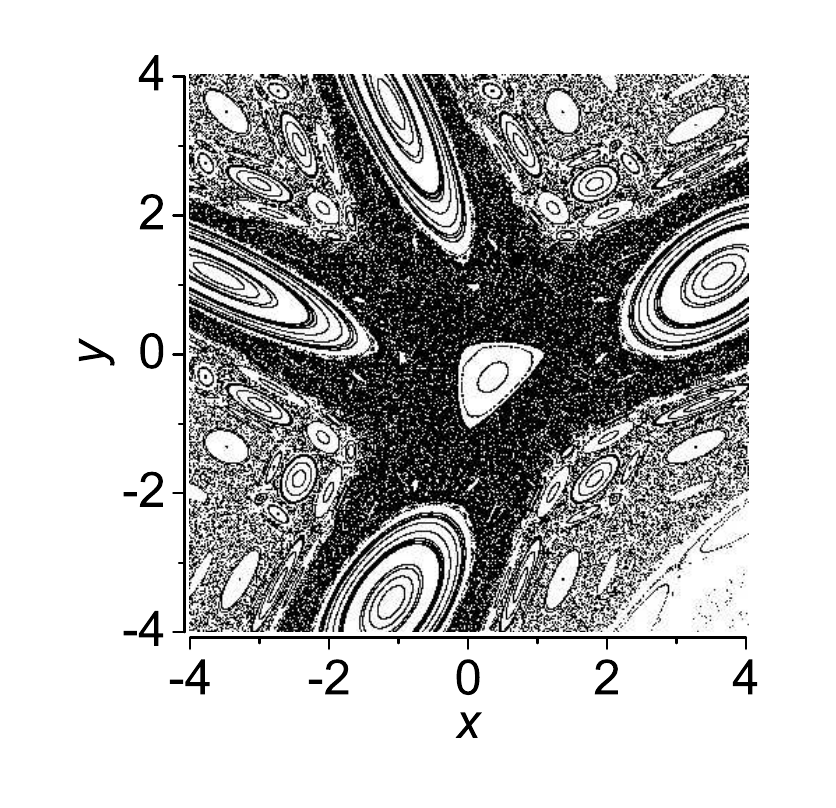}
\caption{\label{fig:T3plus4} Phase space of the Lozi
map for the values $a=0.9$ and $b=-1$. Notice that the scales are
the same as in Figure \ref{fig:T3}.}

\vspace{2cm}
\includegraphics[scale=0.8]{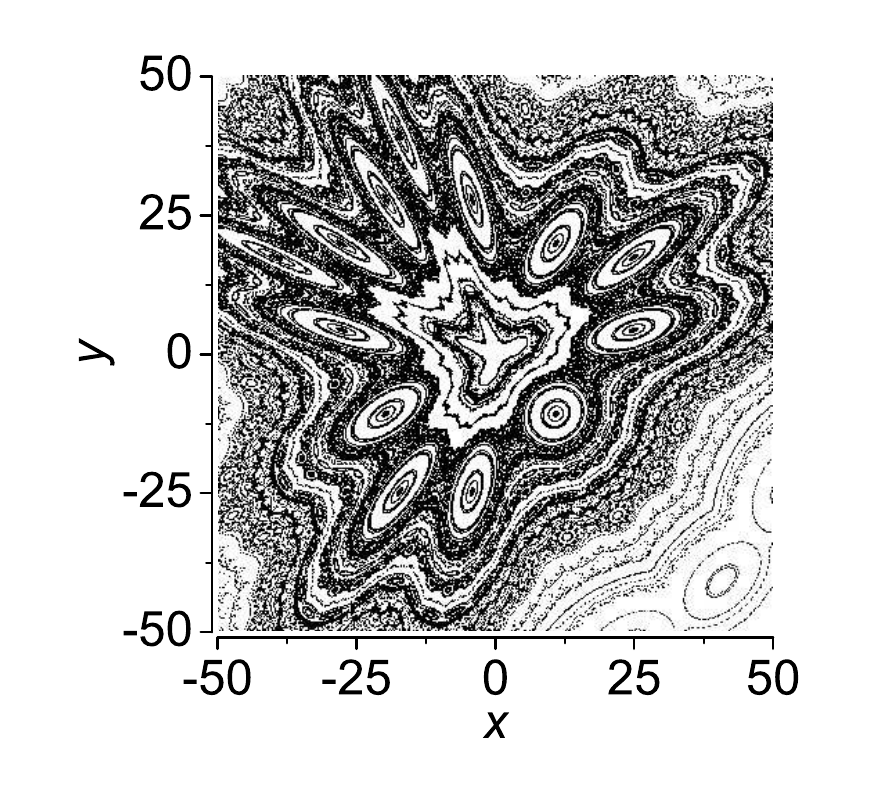}
\caption{\label{fig:T3plus50}  Phase space of the Lozi
map for the values $a=0.9$ and $b=-1$. }

\end{figure}

The phase space shown in Figure \ref{fig:T3} exhibits a mixed structure
of periodic an chaotic trajectories clusters. The complexity of the phase
space is general for $|b|=1$, even for non critical cases as $a= 1$.
Figure \ref{fig:T3plus4} shows the phase space, at the same scale as Figure \ref{fig:T3},
of the neighbouring system $a=0.9,b=-1$. Figure \ref{fig:T3plus50} shows the
intricate phase space structure in a larger scale. It is worth noticing that
for conservative systems ($|b|=1$) trajectories fill up densely any window in phase space.
Thus, to obtain these kind of plots, one has to chose not only a large enough transient but also
an adequate number of initial conditions and iterates to be plotted. Otherwise, the plot will
hardly reveal inner structures.

\section{Final comments} \label{sec:Fin}

In this last section we mention some works related to the Lozi map
which have some bearing with our results.

An interesting study of the dynamics of a H\'enon--Lozi--type map is
carried out in \cite{Grebogi}. The authors introduce a $C^1$ smooth
map which depends on a parameter $\epsilon$. In the limit $\epsilon
\to 0$ their map becomes the Lozi map, and they study which
properties are preserved after the limiting procedure. In
particular, they point out the interesting result that, whereas the
smooth Lozi--like map presents a period--doubling route to chaos,
the genuine Lozi map does not. This observation buttresses the
interpretation advanced in the previous section. Instead of the common infinite sequence of successive period duplications, here we can observe a transition either from period 1 to 2
for $b>1$, or none for $b<1$, before entering the chaotic regime
with the distinctive feature that at the bifurcation points there is
an infinite set of neutrally stable cycles filling a segment, or a sector, in the phase plane.

The theory of border-collision bifurcations provides a classification and a
description of the bifurcations caused by the collision of fixed points with boundaries in piecewise maps. This classification is based on the linearized version of the map around the collision point, usually called the \emph{border-collision normal form map}
\cite{BG}. However, when the normal form is written for the border collisions present in the Lozi map some of the non-degeneracy conditions assumed in the classification theorems do not apply \cite{diB}. In particular, the requirements related to the non-singularity
of the coefficient matrices of the normal form map are not fulfilled. We refer to \cite{diB} for further mathematical details.

Important explicit results concerning systems with stable periodic
orbits embedded in chaos have been reported in  the literature. In
particular, two works \cite{Devaney, Aharonov} study in depth the
area-preserving parameter-free maps
\begin{equation}
x_{n+1}=1-y_n\pm |x_n|, \qquad y_{n+1}=x_n,
\end{equation}
The former corresponds to the plus sign above and the latter to the
minus sign, which is introduced because some calculations are
considerably shorter. These two maps are, except
for a reflection transformation ($y \to -y$), Lozi maps with
$a=\pm 1,b=1$. In \cite{Aharonov} an infinite set of period$-3$
cycles, as the lowest period, are found embedded in a sea of chaos. They
are essentially the ones we have reported in section \ref{sec:T3}.
In \cite{Devaney} the analogue set is for orbits of period$-6$. In
both cases further infinite sets of higher periods are studied too.

In this paper we have revisited the Lozi map with the purpose of
showing that it presents what we have called bisecting bifurcations:
those which are  mediated by an infinite set of neutrally stable
periodic orbits. We have determined explicitly the location of some
of them as well as their cycle elements. We have also provided an
explanation for their existence in  both algebraic and geometric
terms. We deem that particular analyses, like the one we have
carried out for the Lozi map, may help   to enlarge our knowledge
about the dynamics of piecewise continuous maps.

\ack  This work has been partially supported by contracts MCyT/FEDER, Spain
FIS2007-60133 and MICINN (AYA2010-22111-C03-02). VBS thanks Generalitat
Valenciana for financial support. The authors thank Joe Challenger for
interesting comments on the manuscript.

\setcounter{section}{0}

\section*{References}
\begin{thebibliography}{99}

\bibitem{Lozi} Lozi R 1978 Un attracteur \'etrange (?) du type attracteur de H\'enon
\emph{Journal de Physique} \textbf{39} C5--9

\bibitem{henon} H\'{e}non M 1976 A two-dimensional mapping with a strange
attractor \emph{Comm. Math. Phys.} \textbf{50}69--77

\bibitem{Mcz} Misiurewicz M 1980 Strange attractors for the Lozi mappings
\emph{Ann. N.Y. Acad. Sci.} {\bf 357} 348--358

\bibitem{Bap} Baptista D and Severino R 2009 The Basin of attarction of Lozi mappings \emph{Int. J. Bifurcation  Chaos} \textbf{19}
1043--1049

\bibitem{VGH} Botella-Soler V, Oteo J A and Ros J 2009 Dynamics of a
map with a power-law tail
\emph{J Phys A: Math Theor} \textbf{42} 385101--23

\bibitem{Gardini1} Gardini L, Sushko I, Naimzada A. L. 2008 Growing
through chaotic intervals \emph{J Econ Theory} \textbf{143}  541--557

\bibitem{NY}Nusse H E and Yorke J A 1992 Border-collision bifurcations
including ``period two to period three'' for piecewise smooth
systems \emph{Physica D: Nonlinear Phenomena} \textbf{57}
39--57

\bibitem{BG} Banerjee S, Grebogi C 1999 Border collision bifurcations
in two-dimensional piecewise smooth maps \emph{Phys. Rev. E} {\bf 59}
4052--61

\bibitem{diB} Di Bernardo M, Budd C J, Champneys A R, Kowalczyk P  2008 Piecewise-smooth dynamical
systems: theory and applications  \emph{Springer Verlag}

\bibitem{Gardini2} Sushko I, Gardini L, 2010 Degenerate bifurcations
and border collisions in piecewise smooth 1D and 2D maps,
\emph{Int. J. Bifurcation Chaos}, 20, 7, 2045--2070

\bibitem{Yorke} Alligood K T, Sauer T D and Yorke J A 1996 Chaos: an
introduction to dynamical systems (Springer, NY)

\bibitem{Elhadj} Elhadj Z and Sprott J C 2010 A new simple 2-D piecewise linear map
\emph{J. Syst. Sci. Complex.} {bf 23} 379--389

\bibitem{Grebogi} Aziz-Alaoui M A, Robert C and Grebogi C 2001 Dynamics of a
H\'enon-Lozi-type map \emph{Chaos, Solitons and Fractals}
\textbf{12} 2323--2341



\bibitem{Devaney} Devaney R L 1984 A piecewise linear model for the zones of
instability of an area-preserving map \emph{Physica D} \textbf{10}
387--383

\bibitem{Aharonov} Aharonov D, Devaney R L and Elias U 1997 The dynamics of a
piecewise linear map and its smooth
approximation \emph{Int. J. Bifurcation Chaos} \textbf{7} 351--372

\end{thebibliography}
\end{document}